\documentclass[usenatbib, usegraphicx]{mn2e}

\usepackage{amssymb}

\title[Bending instability in galactic discs. 
Advocacy of the linear theory.]
{Bending instability in galactic discs. Advocacy of the linear theory.} 

\author[S.A.~Rodionov and N.Ya.~Sotnikova]
{S.A.~Rodionov$^{1,2}$\thanks{E-mail: astroseger@gmail.com} 
and 
N.Ya.~Sotnikova$^{2,3}$\\
$^{1}$Aix Marseille Universit\'e, CNRS, LAM (Laboratoire d'Astrophysique de
Marseille) UMR 7326, 13388, Marseille, France \\
$^{2}$St. Petersburg State University,
Universitetskij pr.~28, 198504 St. Petersburg, Stary Peterhof, Russia\\
$^{3}$Isaac Newton Institute of Chile, St. Petersburg Branch\\
}

\date{Accepted ???? ??? ??. Received ???? ??? ??; in original form ???? ??? ??}

\pagerange{\pageref{firstpage}--\pageref{lastpage}} \pubyear{201?}

\begin{document}
\label{firstpage}

\maketitle

\begin{abstract}

 We demonstrate that in $N$-body simulations of isolated disc galaxies
there is numerical vertical heating which slowly increases the 
vertical velocity dispersion and the disc thickness. Even for models 
with over a million particles in a disc, this heating can be 
significant. Such an effect is just the same as in numerical experiments 
by \cite{Sellwood2013}. We also show that in a stellar disc, outside a 
boxy/peanut bulge, if it presents, the saturation level of the bending 
instability is rather close to the value predicted by the linear 
theory. We pay attention to the fact that the bending instability 
develops and decays very fast, so it couldn't play any role in 
secular vertical heating. However the bending instability defines the 
minimal value of the ratio between the vertical and radial velocity 
dispersions $\sigma_z / \sigma_R \approx 0.3$ (so indirectly the 
minimal thickness) which could have stellar discs in real galaxies. 
We demonstrate that observations confirm last statement.
\end{abstract}


%----------------------------------------------------------------------------%
\section{Introduction}

 In this article we study the mechanisms of vertical heating of
stellar discs in isolated disc galaxies by numerical simulations. 
Our main interest is to understand the role which the bending
instability plays in real galaxies. We also will pay attention to the
artificial numerical effects in $N$-body models of the disc galaxies.

% This article is the revision of our previous work 
%(\citealp{SR03}, hereafter SR03) where we have studied the mechanisms 
%of vertical heating of a stellar discs in isolated galaxies, and 
%the role of bending instability in this process.

The bending instability (or ``firehose'' instability) for an 
infinitely thin stellar sheet with the 
non zero velocity dispersion was studied for the first time by 
\citet{Toomre1966}. Based on the linear analysis, \citet{Toomre1966} 
found from qaulitative considerations that such a sheet would be 
unstable to bending perturbations if the ratio of the vertical to 
horizontal velocity dispersions, $\sigma_z/\sigma_x$  (velocity 
anisotropy) does not exceed 0.3. In several subsequent studies 
\citep{Kulsrud+1971, PolShukh1977, Araki1985}, this criterion has been 
specified. 
For example, \citet{PolShukh1977} were the first who made an accurate 
analysis of the corresponding dispersion relation for a 
finite-thickness homogeneous layer and found the critical value of 
the velocity anisotropy to be $\approx 0.37$ (bellow this value the 
system is unstable). 
\citet{Araki1985} revised such an analysis for a more realistic model. 
He took into account a non-uniform volume density distribution in 
the $z$-direction and anisotropic Gaussian velocity dispersion. 
Araki's criterion gives $\sigma_z/\sigma_x \approx 0.29$ that is very 
close to the value obtained by Toomre (see discussion by 
\citealp{MerrittSellwood1994}).

 In our previous article (\citealp{SotRod2003}, hereafter SR03) we 
claimed that in numerical models of realistic stellar discs the 
saturation level of the bending instability is actually significantly 
higher than the value ~$0.3 - 0.37$ predicted by the linear theory, 
i.e the disc is unstable even though $\sigma_z/\sigma_R > 0.3$. 
\citet{MerrittSellwood1994} and \citet{Khop+2010} came to the 
similar conclusion. For example, \citet{MerrittSellwood1994} presented 
the results of $N$-body experiments with finite-thickness discs that 
confirmed instability to bending even when the velocity anisotropy was 
larger than the critical value for the instability in an infinite slab. 
In this article we demonstrate that such a  
conclusion was probably wrong, and the saturation level is actually 
very close to the linear criterion. 

 SR03 considered initially thin models and studied processes which 
cause the vertical heating (thickening) of the stellar disc of isolated
disc galaxies. 
Four mechanisms of vertical heating have been revealed. 
\begin{enumerate}
\item
An initial bending instability which develops and 
decays rather fast (well within the first Gyr of the evolution).
\item
An axisymmetric bending instability in the central parts of hot, barless 
models, connected with the formation of the X-shape structure 
in the central parts of models.
\item
A bending instability (buckling) of a bar, which also causes 
the formation of the X-shape structures.
\item
Slow vertical heating which we explain as heating due to vertical
inhomogeneities in the stellar matter.
\end{enumerate}

 Two of these mechanisms (ii and iii) cause the vertical heating of 
a stellar disc up to the value $\sigma_z/\sigma_R \approx 0.8$. 
Both mechanisms lead to the formation of boxy or peanut structures 
in the central part of the model. The backbone of these structures is 
stable families of periodic orbits which bifurcate from the planar 
x1 family in the presence of vertical resonances \citep{Patsis+2002}. 
The existence of vertical resonances is independent of $m=2$ 
perurbations in the disc plane so that X-shape structure may appear in 
almost axisymmetric cases (ii). It should be noted that this X-shape
structure develops asymmetrically in respect with disc plane (see fig. 2
in SR03), so at initial stages it looks like bending of the disc.

 Two points are of the essence.
First, these two mechanisms are probably caused by some type of the 
orbit instability \citep{Skokos+2002, Patsis+2002}, not by the bending 
instability. So, it is expectable that this process is not the subject 
to the linear criterion of the bending instability. 
Second, from the observational point of view a boxy (or peanut) 
structure in the central part of a galaxy is a bulge (or a pseudo 
bulge), not a disc. So if we talk about the disc we should consider 
regions outside this structure.

 In this article we will demonstrate that, 
in the absence of the boxy/peanut structure, the bending instability 
heats the stellar disc up to the value of $\sigma_z/\sigma_R$ not 
higher than $\sim 0.3$, which is predicted by the linear theory. In 
the presence of boxy/peanut structures the above statement is correct 
only for regions outside these structures (in the case of models with 
a bar --- outside the bar). 

 In SR03 we did not discuss regions outside the boxy/peanut structure 
(so outside a bar for models with bars, or outside the X-shaped 
structure for dynamically hot models). But from the data published 
in SR03 one can find that the value of $\sigma_z/\sigma_R$ in a very 
disc is not exceed $\sim 0.35$ (see fig.~5 in SR03), i.e. 
the saturation level of the initial bending instability is close to 
the linear criterion. Here we confirm this result for models with a 
live halo. 

 We also will demonstrate that slow vertical heating 
(the (iv) mechanism), which was explained in SR03 as heating due to 
vertical inhomogeneities in the stellar matter, is a purely numerical 
effect, and for sufficient number of particles there is no heating of 
this kind at all even in the presence of a live halo.

 In section~\ref{s_nummodel} we describe our numerical models, in
section~\ref{s_results} we discuss the 
results of our numerical
simulations and we conclude in section~\ref{s_conc}.

%%%%%%%%%%%%%%%%%%%%%%%%%%%%%%%%%%%%%%%%%%%%%%%%%%%%%%%%%%%%%%%%%%%
\section{The numerical model}
\label{s_nummodel}
 We consider four families of models. 
Each family consists of three models with identical parameters but 
with a different number of particles (i.e. with different spatial 
resolution). All our models include a stellar disc and a dark halo.
Two types of halos are used: a relatively massive one (halo ``A'') and
a relatively light one (halo ``B''). 

 For the first family we model the potential of a halo as static one 
(these models are analoguos to the model 9\_1 from SR03). We denote 
this family as ``AR''. 

 The second family is identical to the first family but all models 
have a live halo. We call this family ``AL''. 

 The models from the third family (``ATL'') are similar to ``AL'' but 
they have an initially thick stellar disc in which we do not expect 
the bending instability to develop. 

 The forth family (``BL'') includes a live halo of the type~B.

 In each family we consider three models with different number of
particles in the disc: 
$2 \cdot 10^5$, $1 \cdot 10^6$ and $5 \cdot 10^6$. 
The models with a live halo have the number of particles in the halo 
$5 \cdot 10^5$, $2.5 \cdot 10^6$ and $5 \cdot 10^6$, 
respectively.

% \footnote{Each number corresponds to increasing 
% number of particles in a disc.}

 We refer to the models under the name of their family, adding 
the number of particle in the disc. For example AR.5M is the model 
from the family AR with five million particles in the disc.

 While constructing the initial equilibrium models for our 
simulations we use the iterative method presented in 
\citet{RodSot2006} and \citet{Rodionov+2009}.

 We use the following density distribution in a stellar disc:

\begin{equation}
\label{eq_star_disc_dens}
\rho_{\rm d}(R,z)=\frac{M_{\rm d}}{4 \pi h^2 z_0}
\cdot \exp\left(-\frac{R}{h}\right)
\cdot {\rm sech}^2\left(
\frac{z}{z_0} \right) \, ,
\end{equation}
where $M_{\rm d}$ is the total disc mass, $h$ is the disc scale length,
$z_0$ is its scale height and $R$ is the cylindrical radius. For all
models discussed here $M_{\rm d} = 5 \cdot 10^{10} M_\odot$ and
$h=3 \; \rm kpc$. The disc in all models except the family ALT has
$z_0 = 0.01 \; \rm kpc$. The initial disc is extremely 
(unrealistically) thin because we are studying how it thickens 
due to the bending instability. 
For the ``thick'' ALT models $z_0 = 0.1 \; \rm kpc$.

 For all models we prepared the disc with the velocity dispersion 
profile $\sigma_R$ as the input (see also \citet{Rodionov+2009} for 
details). The radial profile of $\sigma_R$ together with the density 
distribution in the disc defines the profile of the Toomre parameter 
$Q_\mathrm{T}$ \citep{Toomre1964}. We choose the profile of $\sigma_R$ so that 
the profile of $Q_\mathrm{T}$ has a wide minimum in the region of 
$2-8$ kpc, and the minimal value of $Q_T \sim 1.5$. This value is 
justified by the results of numerical experiments by 
\citet{Khop+2003}. 
Modelling marginally stable stellar discs with various input data 
they demonstrated that $Q_\mathrm{T}$ varies along the radius passing 
through a minimum $Q_\mathrm{T} = 1.2 - 1.6$ just beyond the region 
at the galactocentric distance of $(1-2) \cdot h$, independently of 
the model choice. This value provides marginal in-plane stability of 
the disc and barless models. Starting with such a profile of the 
Toomre parameter, we can supress the bar formation only in the case 
of a rigid halo (as for the AR models). We cannot avoid the bar 
formation in the case of a live halo (as for the AL models).

\noindent
For the AR and AL models we set 
$$
\sigma_R(R) = 80 \cdot \exp\left(-R/9\right) \; {\rm km \, s^{-1}} \, .
$$ 
For the BL models we chose 
$$
\sigma_R(R) = 100 \cdot \exp\left(-R/9\right) \; {\rm km \, s^{-1}} \, .
$$
For the ATL models we have imposed additional constrains. 
We have choosen the profile $\sigma_R$ and the value of $z_0$ 
so that the ratio $\sigma_z/\sigma_R \approx 0.33$ across the disc: 
$$
\sigma_R(R) = 100 \cdot \exp\left(-R/6\right) \; {\rm km \, s^{-1}} \, .
$$

 The dark halo is modelled as truncated NFW halo 
(\citealt{NFW1996,NFW1997})
\begin{equation}
\rho_{\rm h}(r) =  \frac{C_{\rm h} \cdot T(r/r_t)}
{(r/r_h)(1+r/r_h)^2} \, ,
\label{eq_NFW}
\end{equation}

\noindent
with $T(x) = 1/({\rm cosh}(x) + 1/{\rm cosh}(x))$,
where $r_h$ is the halo scale length, $C_h$ is a parameter defining
the mass of the halo, and $r_t$ is a truncated radius. For all models
$r_h=10 \;\rm kpc$, $r_t=15 \;\rm kpc$. 

 For the halo ``A'' $C_h=0.00491212$, which gives the total mass of 
the halo $4.083 \cdot 10^{11} M_\odot$. 
Inside the region of four disc scale lengths 
(inside $12 \;\rm kpc$) the mass of the halo is approximately $3$ 
times greater than the mass of the disc. 
For the halo ``B'' $C_h=0.00163737$, which gives the 
total mass of the halo $1.361 \cdot  10^{11} M_\odot$, so inside 
a sphere of the radius $12 \rm\; kpc$ the mass of the halo is 
approximately equal to the mass of the disc.

We use the following unit system: 
the unit of length is $u_l=1 \;\rm kpc$, 
the unit of the velocity is $u_v=1 \;\rm km/sec$, 
the unit of mass is $u_m = 10^{10}\; M_{\odot}$ and, consequently, 
the unit of time is $u_t \approx0.98 \;\rm Gyr$. 
For simplicity, while converting this time unit into giga years 
we assume that $u_t=1 \;\rm Gyr$.

To simulate the evolution of our models we use the fast $N$-body code 
gyrfalcON \citep{Dehnen2000, Dehnen2002}. The softening length was 
chosen as $\epsilon = 0.005 \;\rm kpc$. We varied the integration 
time step according the rule $0.1 \sqrt{\epsilon/|{\bf a}|}$, where 
${\bf a}$ is the gravitation acceleration. The energy conservation 
for all our models except AL.200K and BL.200K (the two lowest resolution 
models with a live halo) is better than 0.5\%. For AL.200K and 
BL.200K the energy conservation is about 3\%. We made additional 
experiments for these models with a smaller time step, so with the 
better energy conservation, and we did not find any significant 
difference in the final results.

%%%%%%%%%%%%%%%%%%%%%%%%%%%%%%%%%%%%%%%%%%%%%%%%%%%%%%%%%%%%%%%%%%%%%%%%%%%%
\section{Results}
\label{s_results}
 First, we examined the models from the family AR. 
These models have a rather massive rigid halo. 
Inside the region of four disc scale lengths the halo mass is three 
times greater than the disc mass.

%%%%%%%%%%%%%%%%
\begin{figure*}
\begin{center}
\includegraphics[width=16cm]{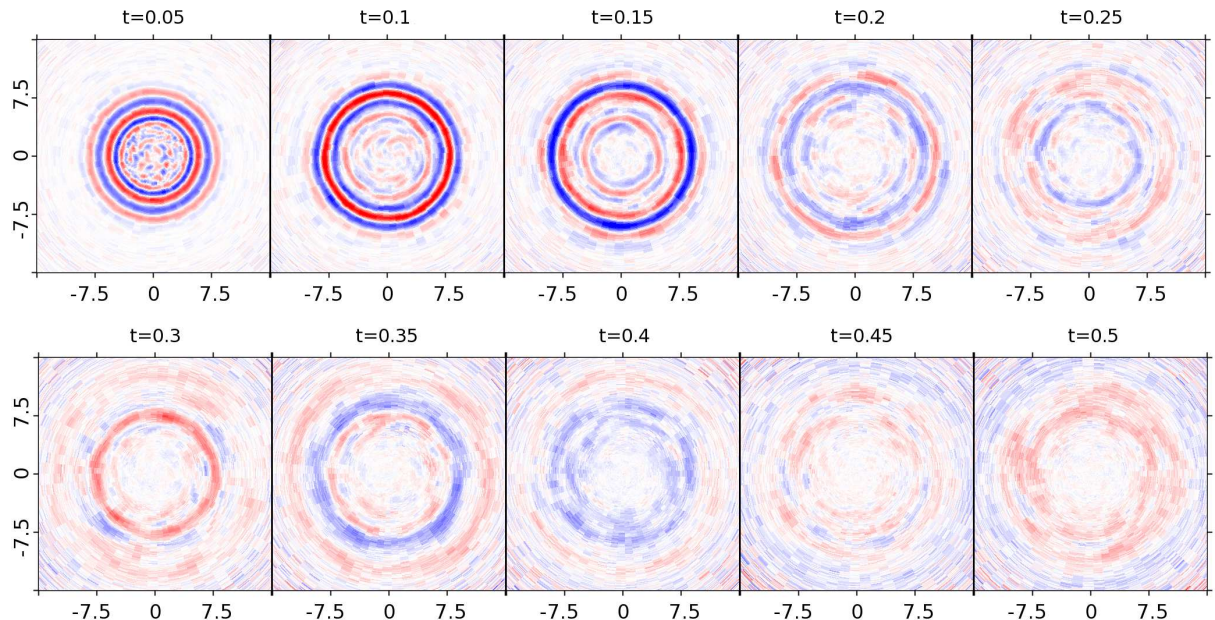}
\end{center}
\caption{Initial stages of the bending instability for the
model AR.5M --- the highest resolution model from the family AR, 
t in Gyr. 
The color intensities correspond to the different values of 
$\bar z$ (the mean value of $z$ coordinate of particles)
in a given area in the disc. Red color means positive values 
of $\bar z$, and blue color means negative values of $\bar z$. 
The range of $\bar z$ is from -0.05 kpc to 0.05 kpc (the most 
intense red color corresponds to $\bar z = 0.05$ kpc, and the most 
intense blue color corresponds to $\bar z = - 0.05$ kpc).}
\label{fig1}
\end{figure*}
%%%%%%%%%%%%%%%%

%%%%%%%%%%%%%%%%
\begin{figure}
\begin{center}
\includegraphics[width=4.1cm, angle=-90]{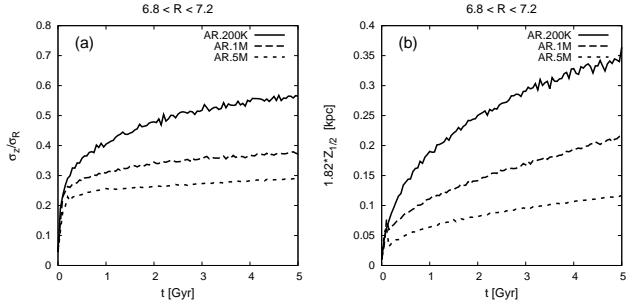}
\end{center}
\caption{The evolution of the ratio $\sigma_z/\sigma_R$ (left panel) 
and of the disc thickness (right panel) for the models from the 
family AR. All values are calculated in an annulus $6.8 < R < 7.2$.}
\label{fig_AR_7kpc}
\end{figure}
%%%%%%%%%%%%%%%%

%%%%%%%%%%%%%%%%
\begin{figure}
\begin{center}
\includegraphics[width=4.1cm, angle=-90]{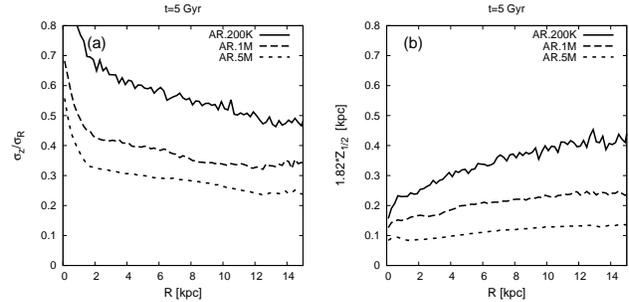}
\end{center}
\caption{The radial profiles of the ratio $\sigma_z/\sigma_R$ and of 
the disc thickness for the models from the family AR at the time 
moment $t=5$~Gyr.}
\label{fig_AR_t5}
\end{figure}
%%%%%%%%%%%%%%%%

 In Fig.~\ref{fig1} we show the evolution of initial bending for 
the highest resolution model from this family ($N = 5 \cdot 10^6$ for
a disc). 
In Fig.~\ref{fig_AR_7kpc} we show the evolution 
of the ratio $\sigma_z/\sigma_R$ and of the disc thickness in the 
area close to $7$ kpc\footnote{We will use the same radius for all our
models, and a radius 7 kpc was chosen in order to be outside the bar 
for the models with a live halo.}. 
Fig.~\ref{fig_AR_t5} demonstrates the radial profiles of these 
parameters at $t=5$~Gyr for all models from the family.
As a measure of the thickness we use the quantity 
$1.82 \cdot z_{1/2}$, where $z_{1/2}$ is the median of $|z|$. 
This quantity is a good estimation of the value of $z_0$ for the 
density distribution~(\ref{eq_star_disc_dens}) and is more robust 
than commonly used dispersion of $z$ (see \citealp{SotRod2006}, 
hereafter SR06). 
For example, for the model AR.5M the thickness of the model
calculated via dispersion of $z$ is almost twice overestimated because 
of a small number of particles with large values of $z$, 
i.e. because of a long tail in the distribution of particles over 
$z$ (SR06).

 The most striking feature in 
Figs.~\ref{fig_AR_7kpc}~and~\ref{fig_AR_t5} is the difference between 
the plots for the same models with different number of particles. 
These models were initially very thin. Therefore the initial 
evolution of these models is rather predictable: disc starts bending 
(see Fig.~\ref{fig1}). The main mode is a ``bell'' mode ($m = 0$). 
It propagates outwards. After approximately $150$~Myr the amplitude 
of bending reaches its maximum and starts decaying. After $1$ Gyr 
bending is almost fully decayed. The thickness and the vertical 
velocity dispersion are growing while bending develops 
(see Fig.~\ref{fig_AR_7kpc}, the very beginning of the evolution). 
We already described this process in SR03 and called it the 
``initial bending''. This initial bending heats up the disc and 
increases the ratio $\sigma_z/\sigma_R$ only to a value which does 
not exceed the value given by the linear criterion of the bending 
instability. Actually, in the case of the AR models, the final value 
of $\sigma_z/\sigma_R$ is even smaller than 0.3 
(see Fig.\ref{fig_AR_7kpc}.a for the model AR.5M).

 During the initial stages of bending the evolution of the AR models 
with different number of particles is similar, but later they start 
diverging. The thickness and the vertical dispersion of the models 
increase with different rates. 
At the moment $t=5$~Gyr the final models have very different 
thickness profiles (see Fig.~\ref{fig_AR_t5}). 
Even for the models with $N = 10^6$ and $N = 5 \cdot 10^6$ 
in a disc the difference in the thickness is about a factor of two. 
We have already described this slow vertical heating in SR03 and 
tried to explain it as heating due to vertical inhomogeneities of the
matter. But this explanation was, at least partly, wrong. 
Despite having an initial disc resistent to bar-like instability 
($Q_\mathrm{T, \, min} \approx 1.5$, rigid halo, 
see \citealp{Khop+2003}), we did not avoid the formation of transient 
small-scale spirals due to noise amplification in our low resolution 
simulations (see fig.~8 in SR03). The spiral activity heats basically 
the in-plane motions. Such a heating effect was noticed for the first 
time by \cite{SellwoodCarlberg1984}. But some authors have worried 
about another effect: discs thicken as spiral activity increases 
(for example, \citealp{Walker+1996,McMillanDehnen2007}). Such a 
relation is rather strange. The enhanced in-plane motions due 
to transient spirals could be redirect into vertical motions only 
by some real or artificial agent. It could be GMCs in real galaxies 
or a separate population of heavy particles in a disc, which was 
not included in numerical models under discussion. 
Actually, such an agent does arise in some numerical models.
In the model 9\_1 from SR03 we 
noticed that the disc thickness is not the same in different parts 
of the model. The disc thickness in the regions where the spiral 
arms are located is smaller than the disc thickness in the interarm 
space. In other words, inhomogeneities in the in-plane 
distribution of stars produce inhomogeneities in the vertical 
distribution of matter and our hypothesis was that such features could
``transfer'' in-plane velocity dispersion into vertical one making 
disc thicker. 

Now we have another explanation. We incline to think that at the
late-stages of the disc evolution slow vertical heating is a pure
numerical effect (a numerical two-body relaxation).
As one can see from Figs.~\ref{fig_AR_7kpc}~and~\ref{fig_AR_t5}, the
greater is the number of particles, the weaker is the vertical heating
effect. But we failed to eliminate
completely this numerical effect because of the limited number
of particles. Even for the model with five million particles
this numerical heating is noticeable. 
\cite{Sellwood2013} has recently resumed the old debate about 
collisional relaxation in very flattened systems. He reminded the 
theoretical arguments by \cite{Rybicki1971} and presented numerical 
results which confirm that two-body scattering in discs is much more 
rapid than that expected in spherical models with the same~$N$. As a result, the 
outcome of simulations will depend on the number of particles used 
in numerical experiments. In the experiments by \cite{Sellwood2013} 
spiral activity lasted too long and heats slowly the in-plane 
motions even for $N = 4 \cdot 10^6$. At the same time, while 
increasing the number of particles up to several million, the disc 
thickenning abruptly lessens.

There is another argument against our old hypothesis. 
In our low resolution models the spiral activity gradually lessens 
and $\sigma_R$ is almost constant at the late stages of evolution 
(see fig.~3 in SR03). In contrast to this in the simulations with 
$N > 10^6$ in a disc the in-plane velocity dispersion $\sigma_R$
continue slowly growing, probably, due to a much clear manifestation
of spirals. Such self-excitation 
behavior has been recently discussed by 
\cite{Henriksen2012}. So one could expect more  efficient vertical
heating in high resolution models, which is actually opposite to
observed results (see figs.~\ref{fig_AR_7kpc}~and~\ref{fig_AR_t5}).
Thus, vertical heating in our simulations is mainly due to the numerical relaxation, 
especially for models with moderate numbers of particles. We cannot 
completely exclude the possibility of out-plane heating due to 
vertical inhomogeneities but the efficiency of such an effect is small.

 Despite the artificial heating revealed in our experiments the final 
value of the ratio $\sigma_z/\sigma_R$ for the highest resolution 
model $A.5M$ is almost everywhere less than $0.3$ 
(see Fig.~\ref{fig_AR_t5}.a). 
Only in the very central region (inside 1~kpc) $\sigma_z/\sigma_R$ is 
relatively large. But in this very central region the model forms a 
subtle X-shaped structure, which is a very efficient source of vertical 
heating (SR03).

%%%%%%%%%%%%%%%%
\begin{figure}
\begin{center}
\includegraphics[width=4.1cm, angle=-90]{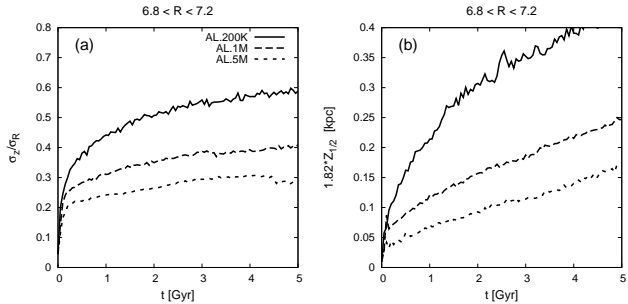}
\end{center}
\caption{The same as in Fig.\ref{fig_AR_7kpc}, but for the family AL.}
\label{fig_AL_7kpc}
\end{figure}
%%%%%%%%%%%%%%%%

%%%%%%%%%%%%%%%%
\begin{figure}
\begin{center}
\includegraphics[width=4.1cm, angle=-90]{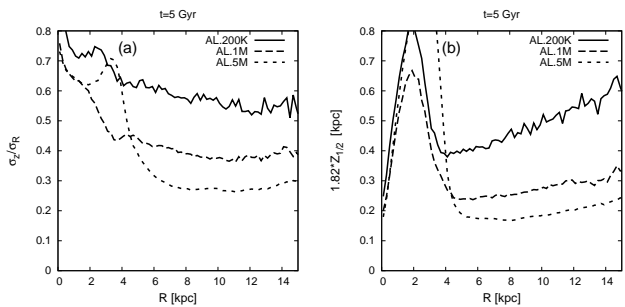}
\end{center}
\caption{The same as in Fig.\ref{fig_AR_t5}, but for the family AL.}
\label{fig_AL_t5}
\end{figure}
%%%%%%%%%%%%%%%%

 Now let us consider the models from the family AL, which are 
completely identical to the models from the family AR, but they 
have a live halo. Fig.~\ref{fig_AL_7kpc} and ~\ref{fig_AL_t5} 
demonstrates the evolution of the ratio $\sigma_z/\sigma_R$ and of 
the disc thickness in the area close to 7 kpc for the models AL.
In Fig.~\ref{fig_AL_t5} the radial profiles of these quantities 
at $t=5$~Gyr are shown. In these models, in contrast with the models 
comprising a rigid halo, there is a bar, which was quite 
expectable \citep{Atan2002}. In all AL models a bar appears after 
approximately a half Gyr of the evolution. The bar grows and then 
slows down due to the angular momentum exchange \citep{Atan2003}. 
While growing, a flat bar starts buckling (i.e. the bending 
instability of the bar develops, see \citealp{Raha+1991}). It results 
in the formation of the boxy/peanut 
structure. This structure can be easily recognized in profiles of 
the disc thickness (see Fig~\ref{fig_AL_t5}.b, inside $R < 6$ kpc). 
As we discussed in SR03 buckling of the bar leads to increase of the 
thickness and of the ratio $\sigma_z/\sigma_R$ in the region of the 
bar. Outside the bar, in the region of the disc, the evolution of 
the AL models is similar to the evolution of AR models, with the 
proviso that the AL models are a little thicker and have a 
slightly larger value of 
$\sigma_z/\sigma_R$ (Figs.~\ref{fig_AL_7kpc}~and~\ref{fig_AL_t5}). 
%But everything which we told about AR models can be applied 
%to AL models if we told about region outside bar.

To summarize:

\begin{enumerate}
\item
the bending instability heats the disc up to the value predicted by 
the linear criterion; 

\item
there is artificial vertical heating which depends on the number of
particles, and even for the models with one and five million 
particles in the disc the difference in the final thickness is 
significant.
\end{enumerate}

The dependence of the effect on the number of particles emploeyd 
supports the idea about two-body relaxation in flattened systems 
\citep{Sellwood2013}.

All our models with a live halo are thicker than the models with a
rigid halo in the region outside a bar (see, for example, high 
resolution models AR.M5 and AL.M5 in 
Figs.~\ref{fig_AR_7kpc}--\ref{fig_AL_t5}). 
It could be due to physical or numerical effects. 
For example some ``physical'' (not just numerical) interaction 
between the disc and the halo could result in disc thickening. 
\cite{Atan2002} showed that a live halo can stimulate bar growth 
by absorbing angular momentum, which a rigid halo cannot. 
\cite{McMillanDehnen2007} were inclined to such an explanation while 
analyzing the greater increase in disc thickness seen in the 
simulations with a live halo compared to that with a rigid halo.
The bar also can influence the areas outside 
itself and make them heat up in the vertical direction. 
Another explanation is more efficient numerical heating in the case 
of a live halo. 
For example, in our highest resolution experiments the mass 
of a particle in the halo is rather large and eight times 
greater than the mass of a particle in the disc. So, halo particles 
can efficiently scatter disc particles. 
Probably, if we took sufficiently large number of particles in a halo 
the model with a live halo would behave similar to the model 
with a rigid halo in the region outside the bar. 
Now we cannot answer which version is right, because
we even cannot claim that the models with rigid halo converge while 
the number of particles increases. 
But we want to underline that the answer to this question 
do not alter our main conclusions about the initial heating due to 
the bending instability.

%%%%%%%%%%%%%%%%
\begin{figure}
\begin{center}
\includegraphics[width=4.1cm, angle=-90]{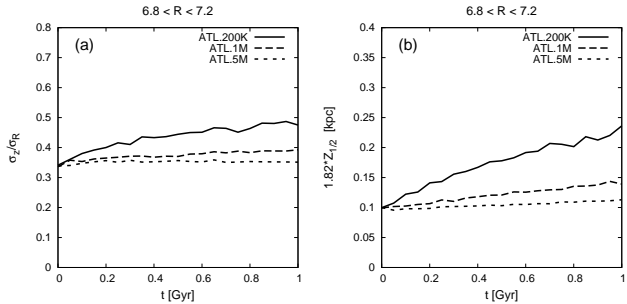}
\end{center}
\caption{The same as in Fig.\ref{fig_AR_7kpc}, but for the family ATL.}
\label{fig_ATL_7kpc}
\end{figure}
%%%%%%%%%%%%%%%% 

 We carried out additional numerical experiments to prove our 
conclusion about numerical heating. We considered the evolution of 
initially relatively thick models (ATL) 
in which we did not expect the bending instability to develop.
In these models the initial ratio $\sigma_z/\sigma_R \approx 0.33$
throughout the disc. 
Fig.~\ref{fig_ATL_7kpc} demonstrates the initial evolution of
the ratio $\sigma_z/\sigma_R$   
and of the disc thickness in the area close to 7 kpc. 
The initial conditions for these models were well beyond the unstable
region of the bending instability, and, as a result, 
these models did not have the stage of initial bending.
Fig.~\ref{fig_ATL_7kpc} shows a pure artificial numerical heating, 
which decreases with increasing the number of particles. 
Starting from almost the same initial conditions 
($\sigma_z / \sigma_R \approx 0.31$ at $t = 0$), 
\cite{Sellwood2013} came to similar results: only for 
$N = 4 \cdot 10^6$ the ratio $\sigma_z / \sigma_R$ did not change 
with time at a reference radius.

%%%%%%%%%%%%%%%%
\begin{figure}
\begin{center}
\includegraphics[width=4.1cm, angle=-90]{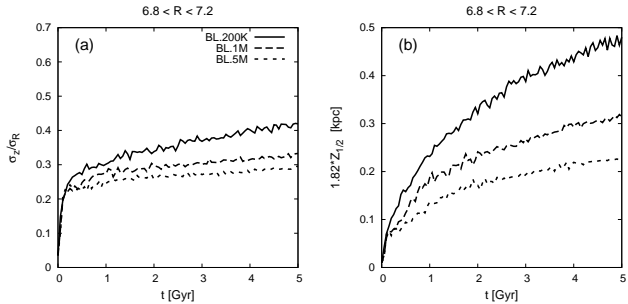}
\end{center}
\caption{
The same as in Fig.\ref{fig_AR_7kpc}, but for the family BL.}
\label{fig_BL_7kpc}
\end{figure}
%%%%%%%%%%%%%%%%

%%%%%%%%%%%%%%%%
\begin{figure}
\begin{center}
\includegraphics[width=4.1cm, angle=-90]{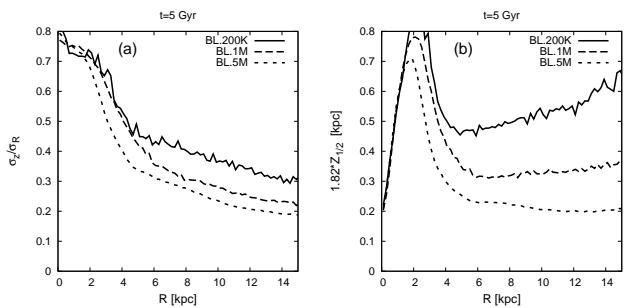}
\end{center}
\caption{
The same as in Fig.\ref{fig_AR_t5}, but for the family BL.}
\label{fig_BL_t5}
\end{figure}
%%%%%%%%%%%%%%%%

 The models, we have just considered had a relatively massive 
dark halo. The models from the family ``BL'' have a less massive 
halo. The halo mass inside the region of four disc scale lengths 
is barely equal to the total disc mass. Fig.~\ref{fig_BL_7kpc} 
demonstrates the evolution of the ratio $\sigma_z/\sigma_R$ and 
of the disc thickness inside the thin ring with radius equal 
to~$7$~kpc. In Fig.~\ref{fig_BL_t5} we show the evolution of radial 
profiles of these quantities at $t=5$~Gyr. The evolution of 
the BL models is similar to the evolution of the AL models. The 
initial bending instability develops. It results in very rapid heating 
of the disc up to the level predicted by the linear criterion. This 
stage is accompanied by the formation of a bar with its subsequent 
buckling and the appearance of the boxy/peanut structure. After 
5~Gyr of the evolution in the area outside the bar, the BL models 
are thicker than the AL models, but they have similar values of 
$\sigma_z/\sigma_R$, because the halo of the BL models is less 
massive. The disc thickness can be estimated via the vertical 
equilibrium condition for an isothermal slab \citep{Spitzer1942}: 
\begin{equation}
\sigma_z^2 = \pi G \Sigma_\mathrm{d} z_0 \, ,
\label{eq_thickness}
\end{equation} 
where $\Sigma_\mathrm{d}$ is the surface density of the slab. This 
relation gives the upper limit for the disc thickness. If there
is a massive dark halo, the disc thickness will be lower for the same
value of $\sigma_z$ (see Eq.~4 in SR06 and 
Eq.~\ref{z0_h} in this paper). 

Again we can see that the models with different numbers of particles 
are significantly different. So, there is numerical heating, however 
in the case of BL family the difference between low and high 
resolution models is less dramatic.
%Probably, it is due to comparable 
%masses of particles in the disc and in the halo so that halo 
%particles scatter disc particles less efficiently.
Values of $\sigma_z/\sigma_R$ for the models BL.1M and BL.5M 
are close (see Figs.\ref{fig_BL_7kpc}.a~and~\ref{fig_BL_t5}.a), 
but the thickness is more visibly different 
(see Figs.\ref{fig_BL_7kpc}.b~and~\ref{fig_BL_t5}.b).

 Similar to previous results, the highest resolution model in the BL 
family (BL.5M model) outside the bar has $\sigma_z/\sigma_R$ that 
corresponds to the linear criterion of the bending instability.

%%%%%%%%%%%%%%%%%%%%%%%%%%%%%%%%%%%%%%%%%%%%%%%%%%%%%%%%%%%%%%%%%%%%%%%%%%%%
\section{Summary}
\label{s_conc}

% We have revised the results of the article SR03 where we
%have studied the mechanisms of vertical heating of stellar discs in
%isolated disc galaxies.

 Modelling how the bending instability develops in $N$-body 
simulations, a few authors 
\cite{MerrittSellwood1994}, SR03, \cite{Khop+2010} proceeded from the 
assumption that the saturation level of the instability was much higher 
than that given by the linear criterium. \cite{MerrittSellwood1994} 
found unstable long-wavelength modes ($m=0,\,1,\,2$) in thin 
finite-thickness discs with realistic density profiles even when 
the ratio $\sigma_z / \sigma_R$ is much less extreme than the critical 
value for the instability in an infinite slab (see their fig.~3). 
Based on the results of a large series of numerical experiments 
SR03 concluded that the saturation level of 
large-scale bending perturbations is a factor of 2 or 3 higher 
than the linear one. \cite{Khop+2010} have been studied the 
non-linear dynamics of the bending instability and the vertical 
structure of a stellar disc embedded into a spherical halo. 
They considered the development of the bending instability to be the
main factor of the disc thickness increase. 

 Slow gradual heating of a disc in the verical direction was 
usually thought as real and triggered by the bending instability not by 
the numerical relaxation because in 
such experiments the increase of $\sigma_R$ eventually ceased when 
the activity of transient small-scale spirals decayed. Moreover, 
suggesting the galaxies to be marginally stable against bending 
perturbations, \cite{Zasov+1991, Zasov+2002} and 
\cite{Khop+2010} tried 
to estimate the contribution of a dark halo into the galaxy mass 
model based only on the value of the disc thicness $z_0/h$ for 
edge-on galaxies (see Eq.~\ref{z0_h}). The critical value for the 
ratio $\sigma_z / \sigma_R$ was taken from the results of $N$-body 
simulations and it was much larger than $\sigma_z / \sigma_R = 0.3$ 
(see fig.~1 in \citealt{Zasov+1991}, 
fig.~5 in \citealt{Zasov+2002} and fig.~12 in \citealt{Khop+2010}).
In SR06 we pointed out that the inclusion in the model a 
compact bulge makes the connection between the relative thickness 
and the halo mass ambiguous (see fig.~6 in SR06). Now we presented 
arguments that in simulations with $N < 10^6$ the value of the 
ratio $\sigma_z / \sigma_R$ was strongly overestimated, while 
athors started from the very ``thin'' initial conditions. 
As \cite{Sellwood2013} have demonstated such an overestimation 
is due to two-body relaxation in flattened rotating systems with 
unsufficient numbers of particles employed.

%We have demonstrated that there is a slow numerical heating in the 
%vertical direction for relatively thin models of stellar discs. 
So, the slow increase of the thickness and of the vertical velocity 
dispersion, which we have observed in the model 9\_1 in SR03, was 
completely due 
to the numerical effect. For a relatively small number of particles in 
the disc ($N=2 \cdot 10^5$) this numerical heating is drastic. Even 
for the models with $5 \cdot 10^6$ particles in the disc this effect 
can not be eliminated. In future, we should be very careful while 
interpreting the results of numerical simulations concerning disc 
thickening or the increase of the vertical velocity dispersion, 
especially while starting from the very thin models. 
%Fortunately, this artificial heating is important only for extremely 
%thin models like those which were considered here. 
%Initially thick models do not demonstrate such strong numerical 
%heating. 

In SR03 we have shown that if we start from a sufficiently thin 
stellar disc the bending instability is developing at the initial stage 
of the disc evolution. This instability develops and decays rather 
fast and results in the increase of the disc thickness and of the 
ratio $\sigma_z/\sigma_R$. Numerical experiments with larger number 
of particles have demonstrated that during the initial bending the 
ratio $\sigma_z/\sigma_R$ increases till the level corresponding to 
the linear criterion of the bending instability. It gives argument that 
this initial bending is caused by the real bending instability 
described by \citet{Toomre1966}. 

Now we can conclude that for our models of stellar discs in isolated 
galaxies in the absence of a separate agent causing out-plane 
scattering (GMCs, nearby external galaxies) there are three mechanisms of vertical 
heating:

\begin{enumerate}
\item The initial bending instability, which develops and decays 
within the first Gyr of the evolution. It heats the disc up to the 
level corresponding to the linear criterion of the bending instability.
\item
The formation of X-shape (boxy/peanut) structure in the central parts
of some barless models.
\item 
The bending instability of the bar (buckling of the bar), which also 
causes the formation of boxy/peanut structures.
\end{enumerate}

It is possible that the last two mechanisms are not associated with
the bending instability but are related to the orbital instability 
\citep{Skokos+2002,Patsis+2002}. Both mechanisms operate in the 
central parts of the galaxy and are responsible for the formation of 
boxy/peanut structures, which perhaps correspond 
to a pseudo bulge in real galaxies. As follows from our simulations, 
the presence of 
boxy/peanut structures does not affect areas outside them. Outside 
the bar the value of $\sigma_z/\sigma_R$ is not greater than that 
predicted by the linear theory of the bending instability, and the 
stellar disc remains very thin.

 If we talk about the disc itself (not about the central regions of
a pseudo bulge) then in our models there is nothing more than 
the classical ``linear'' bending instability which heats the disc up 
to the value $\sigma_z/\sigma_R \approx 0.3$.
This value is rather small, and \citet{Toomre1966} doubted about the 
role of the bending instability in disc heating. Now, we share his 
doubts. But the most important point is that this instability is 
rather fast and does not affect the secular evolution of the stellar 
disc. It develops and decays in less then a half Gyr. 
It cannot produce any long lived out-plane structures in a disc. 
The only role which it could play in real galaxies is to define 
the minimal value of $\sigma_z/\sigma_R$. 
So, in the absence of other mechanisms of vertical heating, 
real galaxies would be marginally stable against the bending 
instability (would have $\sigma_z/\sigma_R \approx 0.3$). 

 One interesting question arises: are there real galaxies,
which marginally stable against the bending instability, or are they 
all completely overheated in the vertical direction due to 
scattering of stars by GMCs or because of interaction with 
nearby satellites.

 The value of $\sigma_z/\sigma_R$ given by the linear criterion of 
the bending instability is rather small, and \citet{Toomre1966} 
emphasized this fact. The ratio of $\sigma_z/\sigma_R$ in the solar 
neighborhood is about 0.5 
(for example, \citealp{DehnenBinney1998,Orlov+2006}) 
and depends weakly on the spectral types of stars. Only for stars of 
spectral classes A and B this ratio is about  0.4 
\citep[table 6]{Mignard2000}.
As to external galaxies, they are rather thick: the typical value of 
$h/z_0$ is less than 5, and for the most thin of them this ratio is 
less then 10 \citep{Mosenkov+2010, Kregel+2002}. 
By combining the equation of equilibrium in the vertical direction
(Eq.~(\ref{eq_thickness})),
the condition of stability in the disc plane \citep{Toomre1964}, and neglecting the
bulge contribution in the mass model we can derive the following
expression
\begin{equation}
\frac{z_0}{h} \approx 0.62 \frac{\sigma_z^2}{\sigma_R^2}
\frac{M_d(4 h)}{M_d(4 h) + M_h(4 h)} Q_\mathrm{T}^2 \, ,
\label{z0_h}
\end{equation}
where $Q_\mathrm{T}$ is Toomre parameter, $M_d(4h)$ is the disc
mass inside the sphere with the radius of four exponential scale
lenght, and $M_h(4h)$ is the halo mass inside $4h$
(see deriving of this equation in SR06).
If we assume that a galaxy has $\sigma_z/\sigma_R = 0.3$,
$Q_\mathrm{T} = 1.4$, and does not have significat amount of dark 
matter then the value of the ratio $h/z_0$ for such a galaxy would be 
$\approx 9$. So, the galaxy, without dark matter, would be as thin as 
the most thinnest galaxies. But with a dark halo the galaxy would be 
thinner than observed galaxies. However, if we increase the value of 
Toomre parameter the resulting value of $h/z_0$ will decrease. 
For example, a galaxy with $\sigma_z/\sigma_R = 0.3$, 
$Q_\mathrm{T} = 2$, and 
$M_d(4h)=M_h(4h)$, would, again, have $h/z_0 \approx 9$. So we can 
conclude that at least the most thinnest galaxies could be marginally 
stable against the bending instability. From the other hand, we expect 
the galaxies with $h/z_0 = 5$ to be on average overheated and have 
$\sigma_z/\sigma_R > 0.3$.

 Until recently, no reliable data about the shape of the velocity 
ellipcoid ($\sigma_z/\sigma_R$) have been available for external 
galaxies. At present, this ratio  is measured directly in several 
galaxies \citep{Gerssen+1997, Gerssen+2000, Shapiro+2003, GerShap2012}. 
There is a clear trend: the later is the type of a galaxy, the 
smaller is the ratio $\sigma_z/\sigma_R$ (see fig.~ 5 in 
\citealp{GerShap2012}). For the early-type galaxies 
$\sigma_z/\sigma_R$ is significantly greater than 0.3 
(for example, $\sigma_z/\sigma_R = 0.75 \pm 0.09$ for Sab galaxy 
NGC~2985). Thus there are strong arguments that real early-type 
galaxies are overheated with respect to the bending instability because 
they have $\sigma_z/\sigma_R$ much greater than 0.3. It means that 
in early-type galaxies some other mechanisms of vertical heating 
operate and they are not included in our simple models of isolated 
disc galaxies. There are at least two candidates for the role of 
such mechanisms: heating by giant molecular clouds and interaction 
with external galaxies or satellites (see, for example, 
\citealp{Sellwood2010} and the discussion in the introduction 
in SR03). 

Surprisingly, for the late-type galaxies the ratio 
$\sigma_z/\sigma_R$ is close to 0.3 ($0.25 \pm 0.20$ for NGC~2280 
and $0.29 \pm 0.12$ for NGC~3810, \citealp{GerShap2012}). It means 
that these galaxies are marginaly stable against the bending 
instability. This result is a bit shocking, because the late-type 
galaxies possess a large amount of gas including the gas in the 
molecular form. For example, for NGC~3810 CO emission is detected 
\citep{Mao+2010}.
But for uknown reasons the scaterring of stars on molecular clouds is 
ineffective in such galaxies.

Another intriguing result is the absence of any trend of $\sigma_z$ 
and $\sigma_z / \sigma_R$ with the surface density 
$\Sigma_\mathrm{H_2}$ of molecular gas, which is usually thought as 
``three-dimensional heating agent''.  At the same time there is a 
correlation between $\sigma_R$ and $\Sigma_\mathrm{H_2}$. If it real, 
this fact poses the question about the role of GMCs in in-plane 
and out-plane scattering of stars.

Summarizing, we can conclude that the answer on the question 
``does the bending instability play any role in the galaxy evolution'' 
is most likely affirmative. The bending instability defines the 
minimal value of $\sigma_z/\sigma_R$ for real galaxies, and there are 
arguments that galaxies, which marginally stable against the bending 
instability, do exist and have $\sigma_z/\sigma_R \approx 0.3$.

%%%%%%%%%%%%%%%%%%%%%%%%%%%%%%%%%%%%%%%%%%%%%%%%%%%%%%%%%%%%%%%%%%%%%%%%%%%%
\section*{Acknowledgments}
We thank the referee, R. Henriksen, for helpful comments.
We also would like to thank E. Athanassoula for useful
discussions.
This work was partially supported by the grant of 
the Russian Foundation for Basic Research 
(grants 11--02--00471a)

%%%%%%%%%%%%%%%%%%%%%%%%%%%%%%%%%%%%%%%%%%%%%%%%%%%%%%%%%%%%%%%%%%%%%%%%%%%%

\label{lastpage}
\end{document}